\documentstyle[aps,multicol,epsf]{revtex}  

\begin{document}


\title{
Anomalies of the infrared-active phonons 
in underdoped YBCO\\
as an evidence for the intra-bilayer 
Josephson effect
}

\author{D.~Munzar$^{1)}$~\cite{address},  
C.~Bernhard$^{1)}$,  
A.~Golnik$^{1)}$~\cite{Aaddress},
J.~Huml{\'\i}{\v c}ek$^{2)}$, and
M.~Cardona$^{1)}$
} 

\address{
1) Max-Planck-Institut f\"ur Festk\"orperforschung, 
Heisenbergstra{\ss}e 1, D-70569 Stuttgart, Germany, EU\\
2)Department of Solid State Physics 
and Laboratory of Thin Films and Nanostructures, 
Faculty of Science, Masaryk University, 
Kotl{\'a}{\v r}sk{\'a} 2, CZ-61137 Brno, Czech Republic 
} 
\date{15.3.99}
\maketitle

\begin{abstract}
The spectra of the far-infrared $c$-axis conductivity 
of underdoped YBCO crystals 
exhibit dramatic changes of some of the phonon peaks 
when going from the normal to the superconducting state. 
We show that the most striking of these anomalies 
can be naturally explained 
by changes of the local fields acting on the ions 
arising from the onset of inter- and intra-bilayer 
Josephson effects. 

\end{abstract}

\begin{multicols}{2}
The essential structural elements 
of the high-$T_{c}$ superconductors 
are the copper-oxygen planes
which host the superconducting condensate.  
Many experiments, and also some theoretical considerations, suggest  
that these planes are only weakly (Josephson) coupled 
along the $c$-direction.  
Studies of the $c$-axis transport \cite{Rapp,Schlenga} 
and those of the microwave absorption \cite{Matsuda,Shibata},  
and the far-infrared $c$-axis conductivity \cite{Tamasaku,Bentum}
revealing Josephson plasma resonances, 
have established that Josephson coupling indeed takes place  
for planes (or pairs of planes) 
separated by insulating layers 
wider than the in-plane lattice constant. 
It is not fully understood 
why the coupling is so weak and it is debated
whether this is related 
to the unconventional ground state of the electronic system 
of the planes causing a charge confinement \cite{Anderson,Emery}
and/or to the properties of the insulating layers. 
In this context, it is of interest to ascertain 
whether the closely-spaced copper-oxygen planes 
of the so-called bilayer compounds, like YBa$_{2}$Cu$_{3}$O$_{y}$, 
are also weakly (Josephson) coupled. 

In this paper we show
that the far-infrared spectra of the $c$-axis conductivity 
of underdoped YBa$_{2}$Cu$_{3}$O$_{y}$ 
with ${6.4\leq y \leq 6.8}$ 
may provide a key for resolving this interesting issue. 
The spectra exhibit, 
beside a spectral gap that shows up already 
at temperatures much higher than $T_{c}$
\cite{Homes1,Homes2,Schutzman,Hauff,Tajima,Bernhard1}, 
two pronounced anomalous features 
\cite{Homes1,Homes2,Schutzman,Hauff,Tajima,Bernhard1}.  
Firstly, 
at low temperatures  
a new broad absorption peak appears   
in the frequency region between $350 {\rm\,cm^{-1}}$ and  
$550 {\rm\,cm^{-1}}$. 
The frequency of its maximum increases with increasing doping; 
for optimally doped samples this feature disappears. 
Secondly, 
at the same time as the peak forms, 
the infrared-active phonons in the frequency region  
between $300 {\rm\,cm^{-1}}$ and $700 {\rm\,cm^{-1}}$
(in particular their strength and frequency) 
are strongly renormalized. 
This effect is most spectacular 
for the oxygen bond-bending mode at $320 {\rm\,cm^{-1}}$, 
which involves the in-phase vibration of the plane oxygens 
against the Y-ion and the chain ions.   
For strongly underdoped 
YBa$_{2}$Cu$_{3}$O$_{6.5}$ 
with $T_{c}\sim 50 {\rm\,K}$, this mode  
loses most of its spectral weight 
and softens by almost $20 {\rm\,cm^{-1}}$. 
Although the additional peak,  
and the related changes of the phonon peaks (phonon anomalies),
start to develop above $T_{c}$, 
there is always 
a sharp increase 
of the peak magnitude below $T_{c}$ \cite{Bernhard2}. 
Similar effects have also been reported  
for several other  
underdoped bilayer-compounds  
(see, e.g., Refs.~\cite{Zetterer,Litvinchuk,Basov}) 
and for hole-doped ladders in Sr$_{14-x}$Ca$_{x}$Cu$_{24}$O$_{41}$
\cite{Osafune}. 

Van der Marel et al.~have suggested \cite{VdMarel2}
that the additional peak around $450 {\rm\,cm^{-1}}$
could be explained 
using a phenomenological model \cite{VdMarel1} 
of the dielectric response of superlattices  
with two superconducting layers (a bilayer) per unit cell.
The model involves two kinds of Josephson junctions: 
inter-bilayer and intra-bilayer. 
As a consequence, 
the model dielectric function exhibits two zero crossings 
corresponding to two longitudinal plasmons: 
the inter-bilayer and the intra-bilayer one. 
In addition, it exhibits also a pole corresponding 
to a transverse optical plasmon.  
Van der Marel et al.~pointed out   
that the additional peak in the spectra of underdoped YBCO 
may just correspond to the latter plasmon. 
Very recently, they have confirmed their suggestion    
by more quantitative considerations 
regarding the doping dependence of the peak position \cite{VdMarel2}. 
The details of the spectacular anomaly
of the $320 {\rm\,cm^{-1}}$ phonon mode, 
however, cannot be explained
within the original form of their model.  

In the following we report a theoretical analysis 
of the additional peak 
and the phonon anomalies. 
We have extended the model of van der Marel et al.~by  
including the four phonons at 280, 320, 560, and $630 {\rm\,cm^{-1}}$  
\cite{Humlicek,Henn}
in such a way that the extended model can account not only 
for the peak but also for the most striking phonon anomalies. 
The important new feature is 
that we take into account local electric fields 
acting on the ions 
participating in the above mentioned phonon modes.  
As we show below, the phonon anomalies  
are then simply due 
to dramatic changes of these local fields 
as the system becomes superconducting. 

Let us briefly introduce the model.  
The dielectric function is written as 
$$
\varepsilon(\omega)=\varepsilon_{1}(\omega)+i\varepsilon_{2}(\omega)=
\varepsilon_{\infty}+{i\over \omega\varepsilon_{0}}
\sum_{n}{\langle j_{n}(\omega)\rangle\over  E(\omega)}\,,
\eqno (1) 
$$
where $\varepsilon_{\infty}$ 
is the interband dielectric function 
at frequencies somewhat above the phonon range, 
$j_{n}$ are the induced currents, 
$\langle \rangle$ means the volume average, 
and $E$ is the average electric field 
along the $c$-axis. 
The following currents have to be taken into account: 
the Josephson current between the planes of a bilayer, 
${j_{bl}=-i\,\omega\,\varepsilon_{0}\,\chi_{bl}\,E_{bl}}$, 
the Josephson current between the bilayers, 
${j_{int}=-i\,\omega\,\varepsilon_{0}\,\chi_{int}\,E_{int}}$, 
the current due to the oxygen bending mode at $320 {\rm\,cm^{-1}}$,
${j_{P}=-i\,\omega\,\varepsilon_{0}\,\chi_{P}\,E_{loc\,P}}$, 
and the current due to the other three infrared-active modes   
involving vibrations of ions located between the bilayers 
(apical oxygens and chain atoms), 
${j_{A}=-i\,\omega\,\varepsilon_{0}\,\chi_{A}\,E_{loc\,A}}$. 
Here 
\end{multicols}
$$
\chi_{bl}=-{\omega_{bl}^{2} \over \omega^{2}}+
            {S_{bl}\omega_{b}^{2}\over 
	    \omega_{b}^{2}-\omega^{2}-i\omega\gamma_{b}}\,\,,\,\,
\chi_{int}=-{\omega_{int}^{2}\over \omega^{2}}+
            {S_{int}\omega_{b}^{2}\over 
	    \omega_{b}^{2}-\omega^{2}-i\omega\gamma_{b}}\,\,,
\eqno (2) 
$$
$$
\chi_{P}={S_{P}\omega_{P}^{2}\over 
	 \omega_{P}^{2}-\omega^{2}-i\omega\gamma_{P}}\,\,,\,\,
\chi_{A}=\sum_{n=1}^{3}\,{S_{n}\omega_{n}^{2}\over 
	 \omega_{n}^{2}-\omega^{2}-i\omega\gamma_{n}}
\eqno (3) 
$$
\begin{multicols}{2}
\noindent 
are the susceptibilities that enter the model. 
The plasma frequencies 
of the intra-bilayer and the inter-bilayer Josephson plasmons 
are denoted as $\omega_{bl}$ and $\omega_{int}$, respectively. 
We do not attribute any physical interpretation 
to the Lorentzian terms in Eq.~(2) that are designed solely 
to represent the featureless residual electronic background 
in the frequency range of interest  
(i.e., from $200 {\rm\,cm^{-1}}$ to $700 {\rm\,cm^{-1}}$)
in a Kramers-Kronig consistent way.  
The response of the phonons is described by
Lorentzian oscillators as usual. 
Further, 
$E_{bl}$ is the average electric field inside a bilayer, 
$E_{int}$ is the average electric field 
between neighbouring bilayers, 
$E_{loc\,P}$ is the local field acting on the plane oxygens,
and $E_{loc\,A}$ is the local field acting on the ions  
located between the bilayers. 
Note that 
by identifying $E_{loc\,P}$   
with the field acting on the plane oxygens
we have neglected 
the contributions of the other ions involved 
in the phonon 
(the Y-ion and the chain ions).  
This seems to be a reasonable approximation, 
since the contribution of the plane oxygens 
to the $320 {\rm\,cm^{-1}}$ mode 
is known to be dominant \cite{Humlicek,Henn}. 

The electric fields $E_{bl}$, $E_{int}$, 
$E_{loc\,P}$, and $E_{loc\,A}$ can be obtained using 
the following set of equations:
$$
E_{bl}=
E'+{\kappa\over \varepsilon_{0}\varepsilon_{\infty}}
-{\alpha\chi_{P}E_{loc\,P}\over \varepsilon_{\infty}}
\,,
\eqno (4) 
$$
$$
E_{int}=
E'
-{\beta\chi_{P}E_{loc\,P}+\gamma\chi_{A} E_{loc\,A}\over 
\varepsilon_{\infty}}
\,,
\eqno (5)
$$
$$
E_{loc\,P}=
E'+{\kappa\over 2\varepsilon_{0}\varepsilon_{\infty}}
\,, 
\eqno (6)
$$
$$
E_{loc\,A}=E'\,,
\eqno (7) 
$$
$$
-i\omega\kappa=j_{int}-j_{bl}\,,
\eqno (8)
$$
$$
E(d_{bl}+d_{int})=E_{bl}d_{bl}+E_{int}d_{int} 
\eqno (9) 
$$
containing two additional variables, $\kappa$ and $E'$.  
The former represents  
the surface charge density   
of the copper-oxygen planes which alternates from 
one plane to the other 
whereas $E'$ is the part of the average internal field $E$
that is not due to the effects of $\kappa$, $\chi_{P}$ and $\chi_{A}$. 
The terms in Eqs.~(4) and (6) containing $\kappa$ 
represent the fields generated by charge fluctuations 
between the planes. 
The terms in Eqs.~(4) and (5) containing 
the phonon susceptibilities  
represent the fields 
generated by the displacements of the ions. 
The values of the numerical factors 
$\alpha$, $\beta$, and $\gamma$ (1.8, 0.8, 1.4)
have been obtained using an electrostatical model \cite{elst.}. 
While the feedback effects of the phonons 
on the electric fields
have to be included in order to obtain 
the observed softening of the oxygen bending mode, 
they are not essential for explaining  
the spectral-weight anomalies. 
Equation~(8) guarantees charge conservation.  
The distances between the planes of a bilayer 
and between the neighbouring bilayers are denoted by 
$d_{bl}$ ($d_{bl}=3.3 {\rm\,{\AA}}$) and 
$d_{int}$ ($d_{int}=8.4 {\rm\,{\AA}}$), respectively.   

Figure 1 (a) shows the experimental spectra 
of the $c$-axis conductivity of YBa$_{2}$Cu$_{3}$O$_{6.5}$ 
with $T_{c}=53 {\rm\,K}$ 
from Ref.~\cite{Bernhard2}.  
Figures 1(b), 1(c), and 1(d) show 
the data for 
(b) $T=300{\rm\,K}$, (c) $T=75{\rm\,K}$ 
and (d) $T=4{\rm\,K}$ 
together with the fits obtained 
by using the model explained above.

The values of the parameters used are summarized in Table 1. 
Those used in computing the room-temperature spectrum 
have been obtained by fitting 
the measured complex dielectric function from Ref.~\cite{Bernhard2} 
(with $\omega_{bl}=0.0$ and $\omega_{int}=0.0$). 
Those used in calculating the $4 {\rm\,K}$ spectrum  
have also been obtained by fitting the data, 
except for $\varepsilon_{\infty}$, $\omega_{P}$,  
and the oscillator strengths of the phonons 
($S_{P}$, $S_{1}$, $S_{2}$, $S_{3}$)
which have been fixed at the room-temperature values.  
The appearance of the additional peak and the anomalies 
already at temperatures higher than $T_{c}$ 
may be caused by pairing fluctuations within the bilayers. 
Motivated by this idea, we have fitted the 75K spectra 
in the same way as the 4K ones allowing only the upper 
plasma frequency ($\omega_{bl}$) to acquire a nonzero value.  
We shall comment on this point below. 
Note that the low-temperature value  
of $\omega_{int}$ ($220 {\rm\,cm^{-1}}$)
is rather close to the one 
obtained from reflectance measurements 
($204 {\rm\,cm^{-1}}$ in Ref.~\cite{Homes2}) 
and that the screened value of $\omega_{bl}$ 
($\sim 500 {\rm\,cm^{-1}}$)
falls into the frequency region 
of a broad peak in the loss function   
(${\rm Im}(1/\varepsilon)$) \cite{Bernhard2},  
which is a signature of a longitudinal excitation. 
\end{multicols} 
\begin{figure*}
        \parbox{17.5 cm}{
        \leavevmode  
        \hbox{%
        \epsfysize = 7.8 cm
        \epsffile{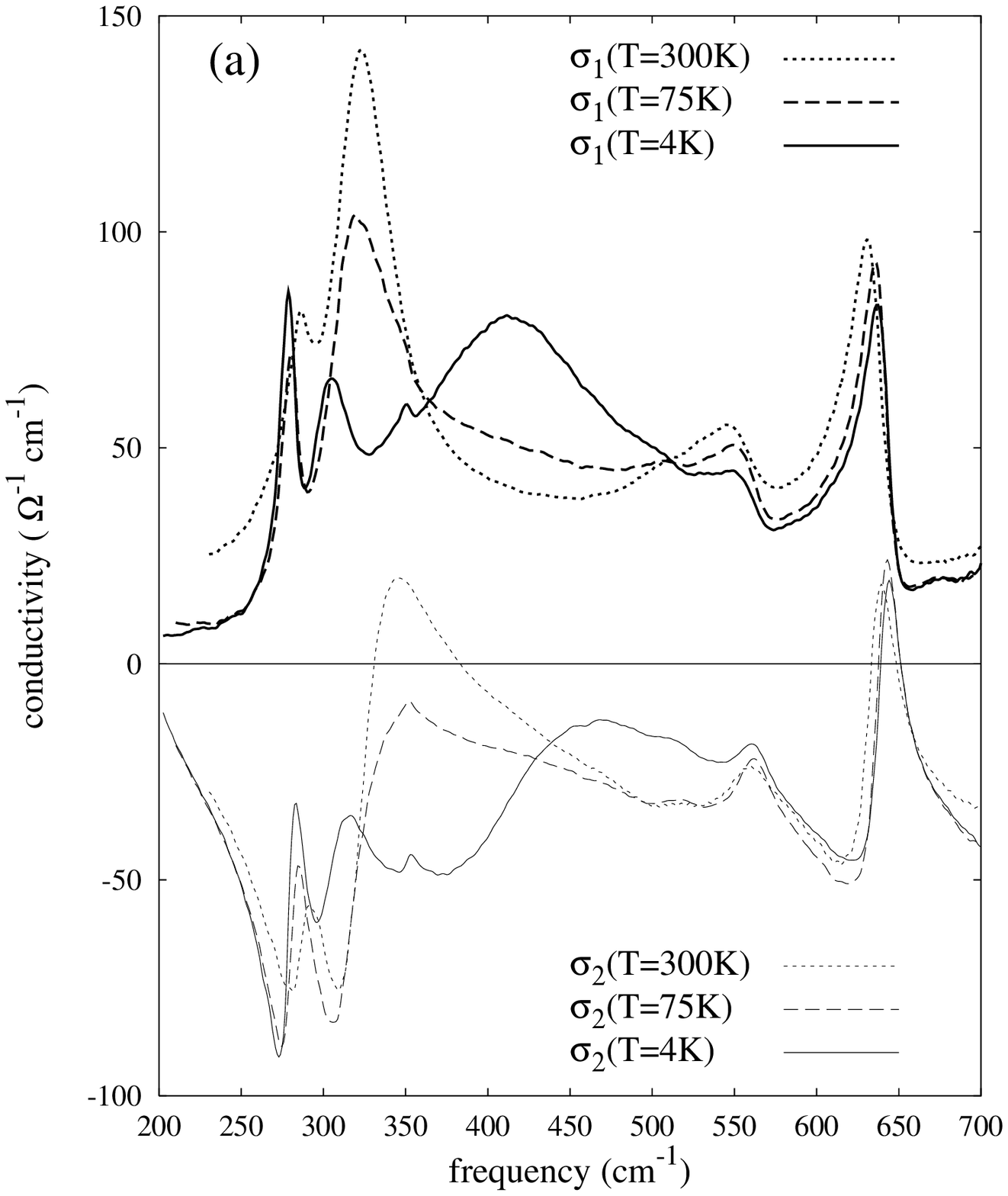} 
        \epsfysize = 7.8 cm
        \epsffile{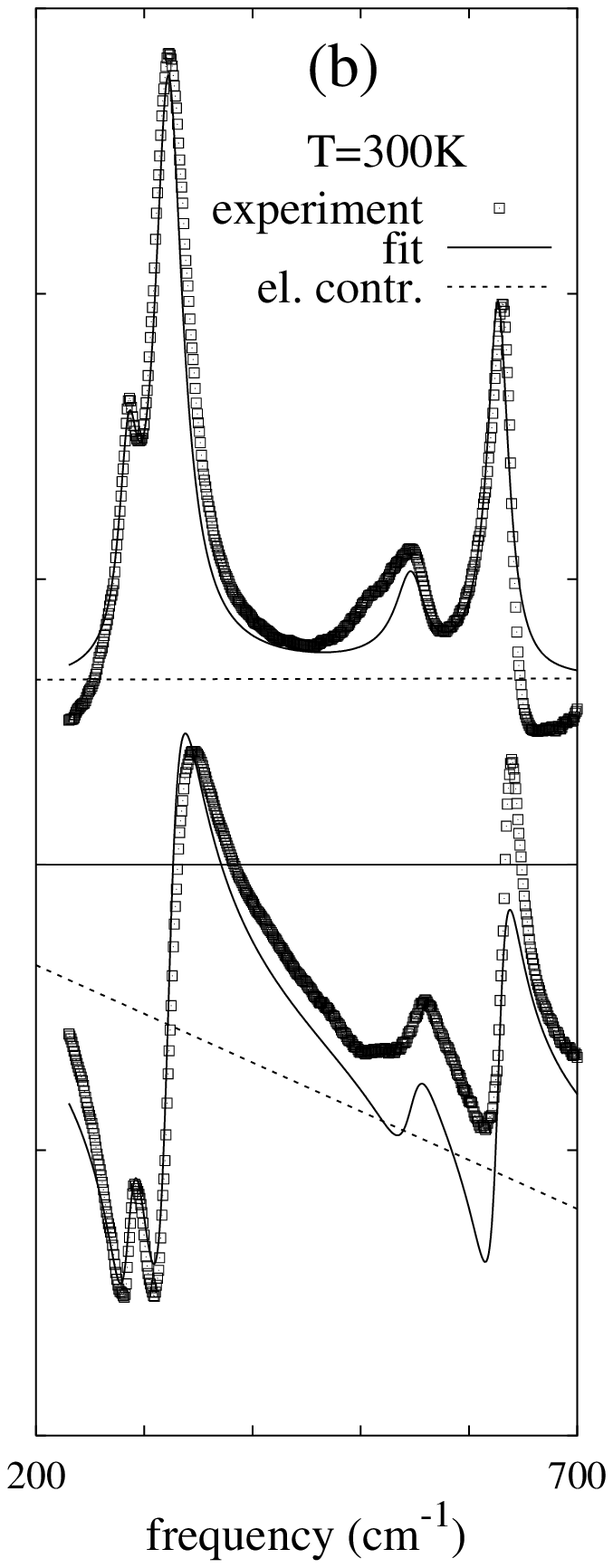} 
        \epsfysize = 7.8 cm
        \epsffile{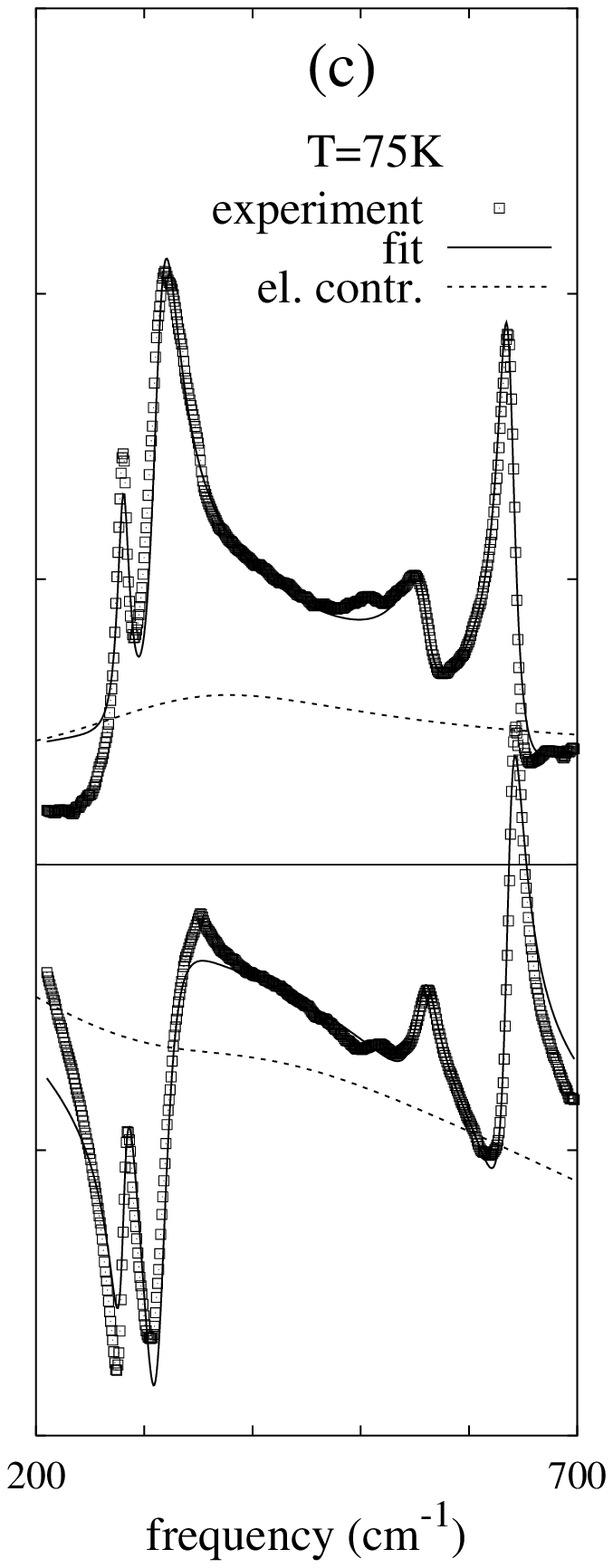} 
        \epsfysize = 7.8 cm
        \epsfbox{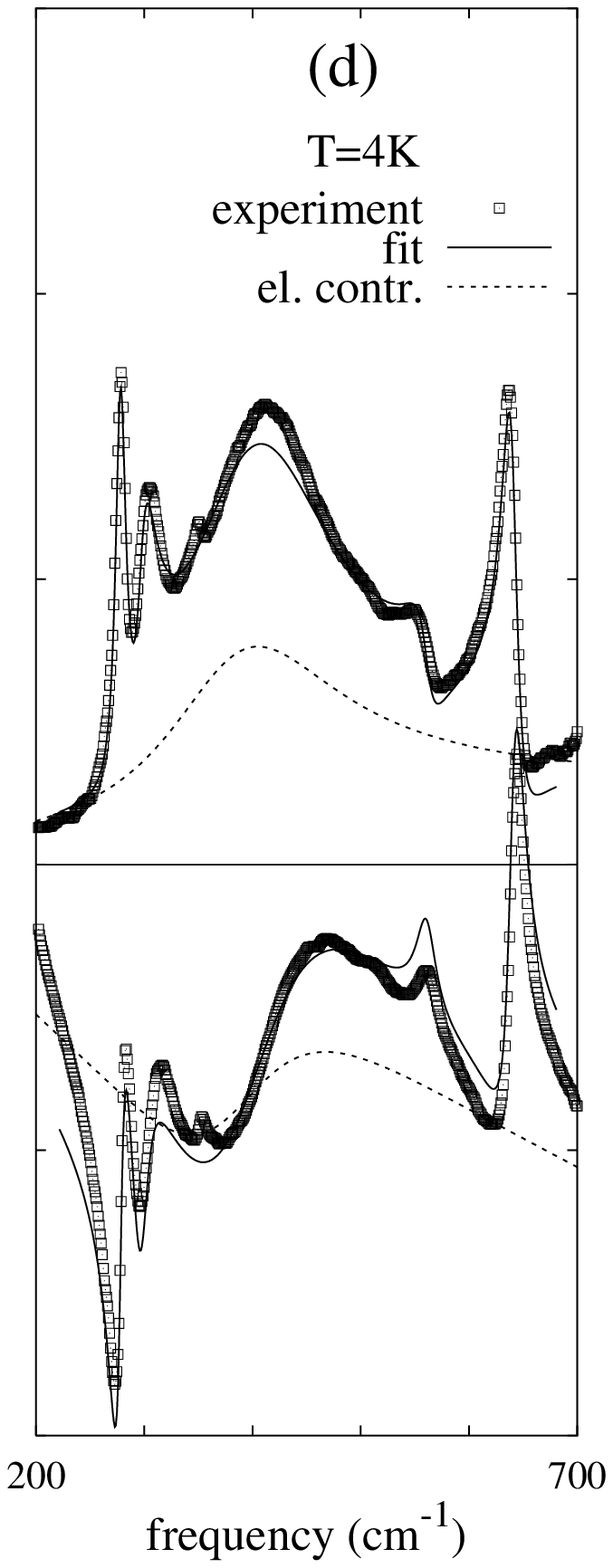} } }
\vskip 0.2cm 
\caption{
(a) Experimental spectra of the $c$-axis conductivity, 
$\sigma=\sigma_{1}+i\sigma_{2}$,
of YBa$_{2}$Cu$_{3}$O$_{6.5}$ with $T_{c}=53 {\rm\,K}$
from Ref.~[15]. 
Experimental data together with the fits 
obtained by using the present model 
for (b) $T=300{\rm\,K}$, (c) $T=75{\rm\,K}$, 
and (d) $T=4{\rm\,K}$.  
The dotted lines represent
the electronic contributions.
}
\end{figure*}
\begin{table*}
\caption{Values of the parameters used in the 
present computation. The temperatures are given in K, 
the frequencies and the broadening parameters 
in ${\rm\,cm^{-1}}$. 
}
\begin{tabular}{ccccccccccccccccccccc}
{$y$}                        &\multicolumn{1}{c} 
{$ T$}                       &\multicolumn{1}{c} 
{$\varepsilon_{\infty}$ }    &\multicolumn{1}{c} 
{$\omega_{bl}$ }             &\multicolumn{1}{c} 
{$\omega_{int}$ }            &\multicolumn{1}{c} 
{$S_{bl}$ }                  &\multicolumn{1}{c} 
{$S_{int}$ }                 &\multicolumn{1}{c} 
{$\omega_{b}$ }              &\multicolumn{1}{c} 
{$\gamma_{b}$ }              &\multicolumn{1}{c} 
{$S_{P}$ }                   &\multicolumn{1}{c} 
{$S_{1}$ }                   &\multicolumn{1}{c} 
{$S_{2}$ }                   &\multicolumn{1}{c} 
{$S_{3}$ }                   &\multicolumn{1}{c} 
{$\omega_{P}$ }              &\multicolumn{1}{c} 
{$\omega_{1}$ }              &\multicolumn{1}{c} 
{$\omega_{2}$ }              &\multicolumn{1}{c} 
{$\omega_{3}$ }              &\multicolumn{1}{c} 
{$\gamma_{P}$ }              &\multicolumn{1}{c} 
{$\gamma_{1}$ }              &\multicolumn{1}{c} 
{$\gamma_{2}$ }              &\multicolumn{1}{c} 
{$\gamma_{3}$ } \\ 
\tableline 
6.5  & 300 & 5.15 & 0 & 0 & 1280 & 1000 & 680 & 250000 &
1.3 & 0.056 &
0.087 & 0.32 & 390 & 290 & 553 & 647 & 21 & 22 & 29 & 16 \\
6.5  &  75 & 5.15 & 994 & 0 & 450 & 170 & 810 & 140000 & 
1.3 & 0.056 &
0.087 & 0.32 & 390 & 285 & 563 & 655 & 9 & 14 & 20 & 7 \\
6.5  &  4 & 5.15 & 1205 & 220 & 8.2 & 2.7 & 6000 & 180000 & 
1.3 & 0.056 &
0.087 & 0.32 & 390 & 284 & 564 & 656 & 9 & 12 & 15 & 3 \\
\tableline
6.8  & 260 & 5.15 & 0 & 0 & 1700 & 1700 & 700 & 220000 & 
1.3 & 0.056 &
0.200 & 0.30 & 390 & 296 & 582 & 631 & 13 & 20 & 15 & 25 \\
6.8  &  4 & 5.15 & 1779 & 450 & 13.8 & 4.8 & 2400 & 15000 & 
1.3 & 0.056 &
0.200 & 0.30 & 390 & 288 & 589 & 646 & 2.5 & 16 & 25 & 20 \\
\tableline 
6.4  & 250 & 5.15 & 0 & 0 & 310 & 240 & 860 & 120000 & 
1.3 & 0.060 & 
0.100 & 0.42 & 400 & 286 & 553 & 650 & 34 & 24 & 30 & 15 \\
6.4  &  4 & 5.15 & 949 & 150 & 560 & 480 & 520 & 120000 & 
1.3 & 0.060 & 
0.100 & 0.42 & 400 & 280 & 556 & 654 & 42 & 24 & 20 & 5 \\
\end{tabular}
\vskip 0.2 cm
\end{table*} 
\begin{multicols}{2}
The values of the phonon frequencies  
are somewhat different from those 
which would result from a usual fit of the data
(such as in Refs.~\cite{Homes1,Homes2,Schutzman,Hauff,Tajima,Bernhard1,Henn}). 
This is because the susceptibilities of Eq.~(3) 
represent response functions 
with respect to the local fields 
instead of the average field.  
In the absence of interlayer currents, 
the input frequencies would correspond to the LO-frequencies 
while the frequencies 
renormalized according to Eqs.~(4)-(9)
would correspond to the TO-ones. 
Our input frequency
of the oxygen bending mode ($390 {\rm\,cm^{-1}}$) 
is close to the measured LO frequency \cite{Humlicek}.  

It can be seen in Fig.~1 
that the model is capable of providing 
a good fit of both the normal- 
and the superconducting-state data  
without changing the oscillator strengths of the phonons
and without any change 
of the input frequency of the oxygen bending mode.  
It reproduces succesfully: 
(i) the appearance of the additional peak, 
its position, broadening and magnitude; 
(ii) the loss of the spectral weight  
of the peak corresponding to the oxygen-bending mode 
and the pronounced softening of this mode; 
(iii) the loss of the spectral weight 
of the peaks corresponding to the apical oxygen modes
at $550 {\rm\,cm^{-1}}$ and $630 {\rm\,cm^{-1}}$
and the increase of their assymetry. 
The intrinsic frequencies of the latter modes 
have to be slightly increased 
in order to reproduce the noticeable hardening of these modes. 
The dotted lines in Figures 1(b),1(c) and 1(d) represent the results 
obtained after omitting the phonons in the fitted expresions 
($S_{P}=S_{1}=S_{2}=S_{3}=0.0$). 
It appears 
that the plasmon peak collects 
the lost part 
of the normal-state spectral weight of the phonons.
This, however, only accounts for a part 
of its spectral weight.  
In the absence of the phonons 
and for small values of the residual conductivities 
the spectral weight of the ${\delta-{\rm peak}}$ at $\omega=0.0$ 
is 
${S_{\delta}=(\pi/2)\varepsilon_{0}
(d_{bl}+d_{int})\omega_{bl}^{2}\omega_{int}^{2}/
(d_{bl}\omega_{int}^{2}+d_{int}\omega_{bl}^{2})}$
and the spectral weight of the additional peak 
is 
$S_{p}=(\pi/2)\varepsilon_{0}
(d_{bl}d_{int}/(d_{bl}+d_{int}))
(\omega_{bl}^{2}-\omega_{int}^{2})^{2}/
(d_{bl}\omega_{int}^{2}+d_{int}\omega_{bl}^{2})$. 
For the values of the two plasma frequencies 
given in Table 1 we obtain ${S_{\delta}=1700\,\Omega^{-1}{\rm cm^{-2}}}$
and ${S_{p}=10 000\,\Omega^{-1}{\rm cm^{-2}}}$. 
Note that both $S_{\delta}$ and $S_{p}$ 
belong to the spectral weight of the 
superconducting condensate. 
This should be taken into account 
in discussing the sum rules 
as, e.g., in Ref.~\cite{Basov2}. 

Our model allows us to explain  
the decrease of the spectral weight of the phonons 
when going from the normal to the superconducting state 
by using rather simple qualitative arguments.  
Neglecting the feedback effects of the phonons
and the residual electronic background 
(represented by the Lorentzians in Eq.~(2)),
the electric fields  
$E_{bl}$ and $E_{int}$ are given 
by: 
\end{multicols}
$$
E_{bl}=
{(d_{bl}+d_{int})
\varepsilon_{int}\over 
d_{bl}\varepsilon_{int}+d_{int}\varepsilon_{bl}}E\,\,,\,\,
E_{int}=
{(d_{bl}+d_{int})
\varepsilon_{bl}\over 
d_{bl}\varepsilon_{int}+d_{int}\varepsilon_{bl}}E\,, 
\eqno (10)
$$
\begin{multicols}{2}
where 
${\varepsilon_{bl}=
\varepsilon_{\infty}-\omega_{bl}^{2}/\omega^{2}}$
and 
${\varepsilon_{int}=
\varepsilon_{\infty}-\omega_{int}^{2}/\omega^{2}}$.  
The low-temperature spectra 
of $\varepsilon_{bl}$ and $\varepsilon_{int}$ 
are shown as the solid lines in Fig.~2.      
In the frequency range of the oxygen bending mode, 
$\varepsilon_{bl}$ and $\varepsilon_{int}$  
have opposite signs and similar magnitudes 
and the same holds for
$E_{int}$ and $E_{bl}$. 
As a consequence,  
the local field acting on the plane oxygens, 
which equals the average of the two fields 
$E_{int}$ and $E_{bl}$ 
(cf.~Eqs.~(4), (5), and (6)), 
can become rather small.   
The frequency range of the modes of the apical oxygen 
is close to the zero crossing of $\varepsilon_{bl}$. 
Consequently, the local field acting on the apical oxygens,  
$E_{int}$, is rather 
small in this frequency region.  
It is the decrease of the local fields 
when going from the normal to the superconducting state 
which is responsible for the spectral weight anomalies. 
The room- and low-temperature spectra of $E_{loc\,P}$ shown 
in Fig.~2 illustrate the above considerations. 
\begin{figure}
\epsfxsize = 8.6 cm
\epsfbox{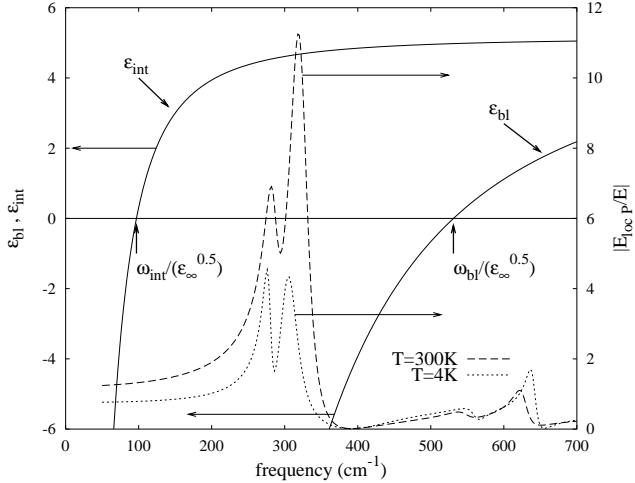}
\vskip 0.2 cm
\caption{
The approximate dielectric functions 
of the intra-bilayer 
and the inter-bilayer regions, 
$\varepsilon_{bl}$ and $\varepsilon_{int}$ defined in the text 
(solid lines). 
The room- and low-temperature spectra 
of the local field $E_{loc\,P}$ 
acting on the plane oxygens (dashed and dotted lines).
}
\end{figure}
Our model is also capable of explaining 
the experimentally observed doping dependence 
of the additional peak 
and the anomaly of the oxygen bending mode. 
As the doping increases  
the peak shifts towards higher frequencies 
and it becomes broader and less pronounced 
(see Fig.~10 of Ref.~\cite{Homes2} and Fig.~3 of Ref.~\cite{Schutzman}).  
Both these trends can be easily understood. 
The first one is due to the progressive increase
of the plasma frequencies with hole doping, which  
is related 
both to the increase of the condensate density 
and to the reduction of the charge confinement \cite{Bernhard1}. 
The second one is due to the fact 
that the broadening is proportional to the 
residual backround conductivities which increase
with increasing doping.  
In addition,  
the size of the spectral gap ($2\Delta_{max}$)
decreases with hole doping \cite{Bernhard1} 
and eventually falls below the energy of the transverse 
optical plasmon around optimum doping. 
The phonon anomaly appears 
in the same range of doping as the additional peak. 
For $y$ around 6.8,  
the spectral weight 
from the high-frequency side of the phonon peak  
moves into the additional peak 
(see Figs.~10 (a) and (b) of Ref.~\cite{Homes2}) 
as the temperature is lowered.  
For $y$ around 6.6 we find the most pronounced anomaly
(see the experimental data 
of Fig.~1). 
For even lower doping levels  
the additional peak and the phonon  
merge together forming  
a single highly-assymetric structure 
(see Fig.~3 (d) of Ref.~\cite{Schutzman}). 
These trends can be understood 
using arguments 
similar to those presented above
(see the discussion related to Fig.~2) 
and can be well reproduced using the model.
This is demonstrated in Fig.~3 which displays 
the experimental spectra 
of the $c$-axis conductivity of YBa$_{2}$Cu$_{3}$O$_{y}$ 
with (a) $y=6.4$ ($T_{c}=25 {\rm\,K}$) and 
     (b) $y=6.8$ ($T_{c}=80 {\rm\,K}$) 
from Ref.~\cite{Bernhard2}  
together with the fits (Fig.~3(c) and 3(d), respectively).

We discuss next 
the peculiar temperature dependence 
of the additional peak and the phonon anomalies. 
The proximity of the onset temperature of the anomalies 
and the onset temperature of the spin gap ($T^{*}$)
observed in nuclear magnetic resonance experiments 
has provoked several speculations \cite{Litvinchuk,Hauff} 
that the anomalies are due to the coupling of the phonons
to spin excitations. 
From the fact that we are able to fit the data 
for temperatures between $T_{c}$ and $T^{*}$  
(see Fig.~1(c)) we infer 
that in this temperature-range 
the intra-bilayer plasmon 
is already developed. 
This suggests 
that many of the electronic and possibly also structural 
anomalies starting 
below $T^{*}$ (see, e.g., Ref.~\cite{Mihailovic}) are caused by 
pairing fluctuations within the bilayers. 
 
In summary, we have extended the phenomenological model 
of Van der Marel et al.~involving 
inter-bilayer and intra-bilayer Josephson junctions 
by including phonons and local field effects. 
\begin{figure}
        \parbox{8.6cm}{
        \leavevmode  
        \hbox{%
        \epsfysize = 4.4 cm
        \epsffile{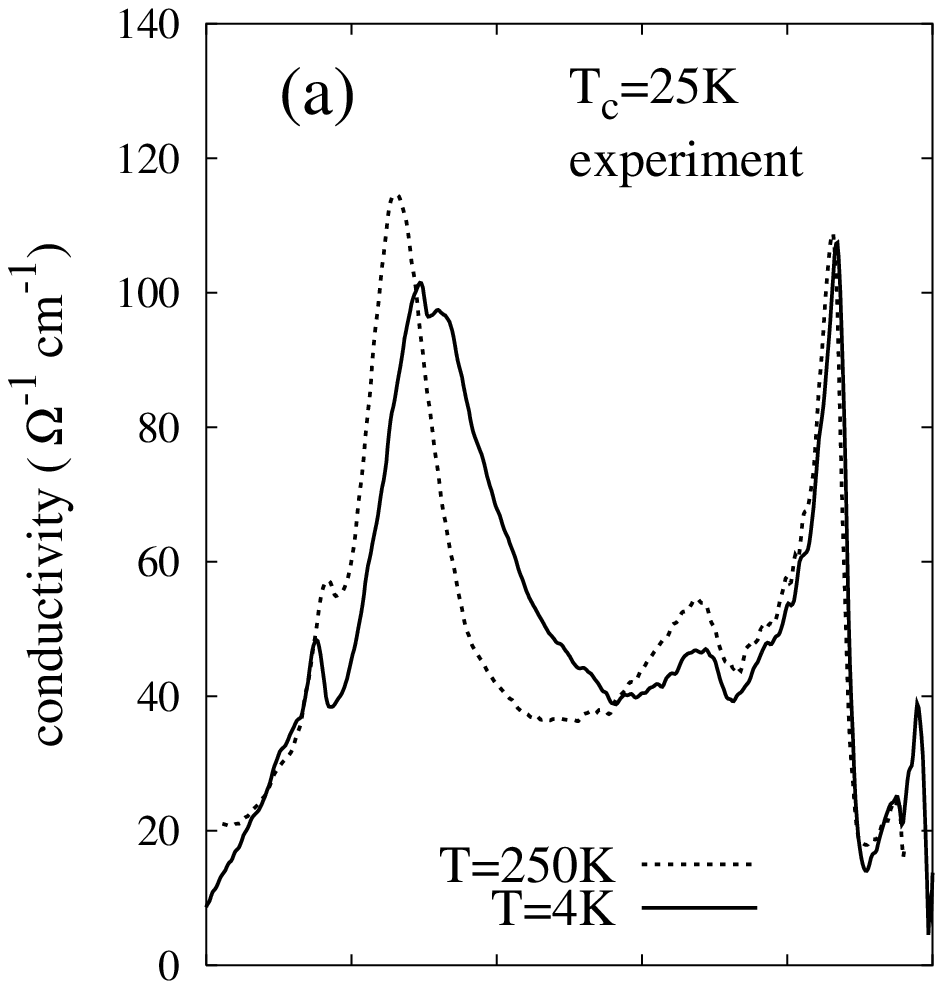} 
        \epsfysize = 4.4 cm
        \epsffile{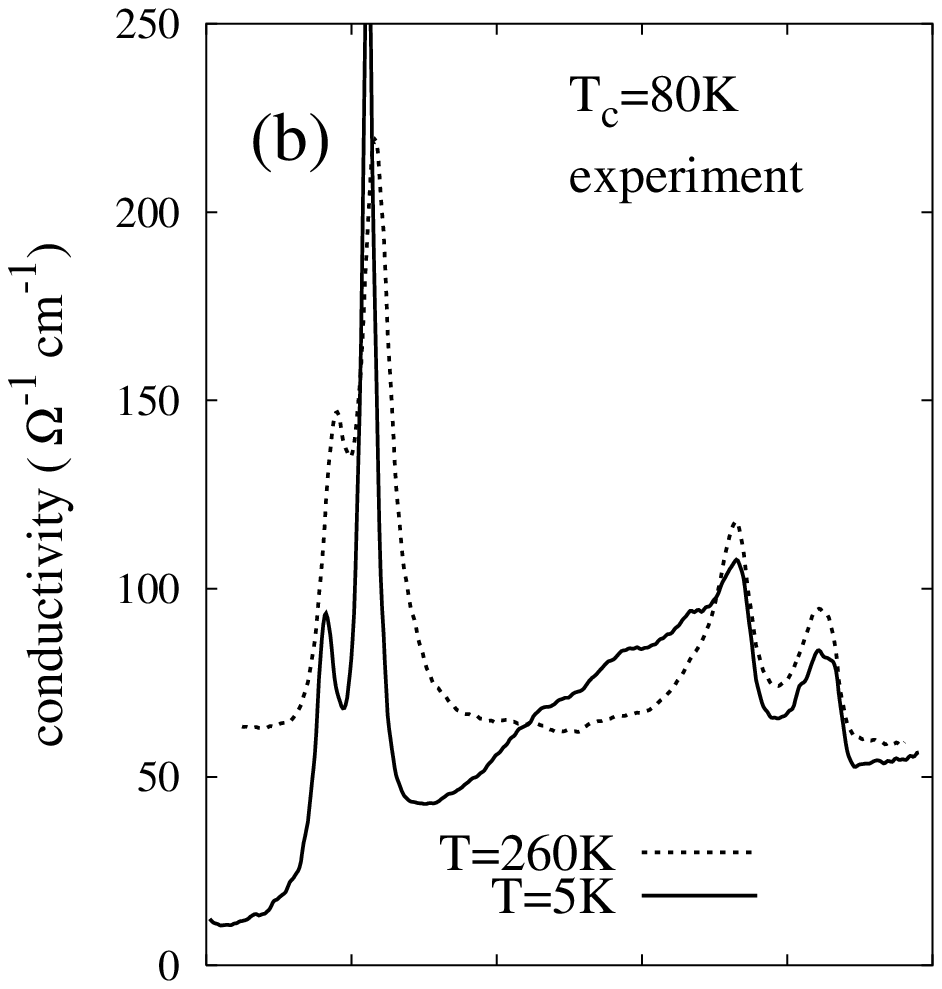} }
        \hbox{%
        \epsfysize = 4.4 cm
        \epsffile{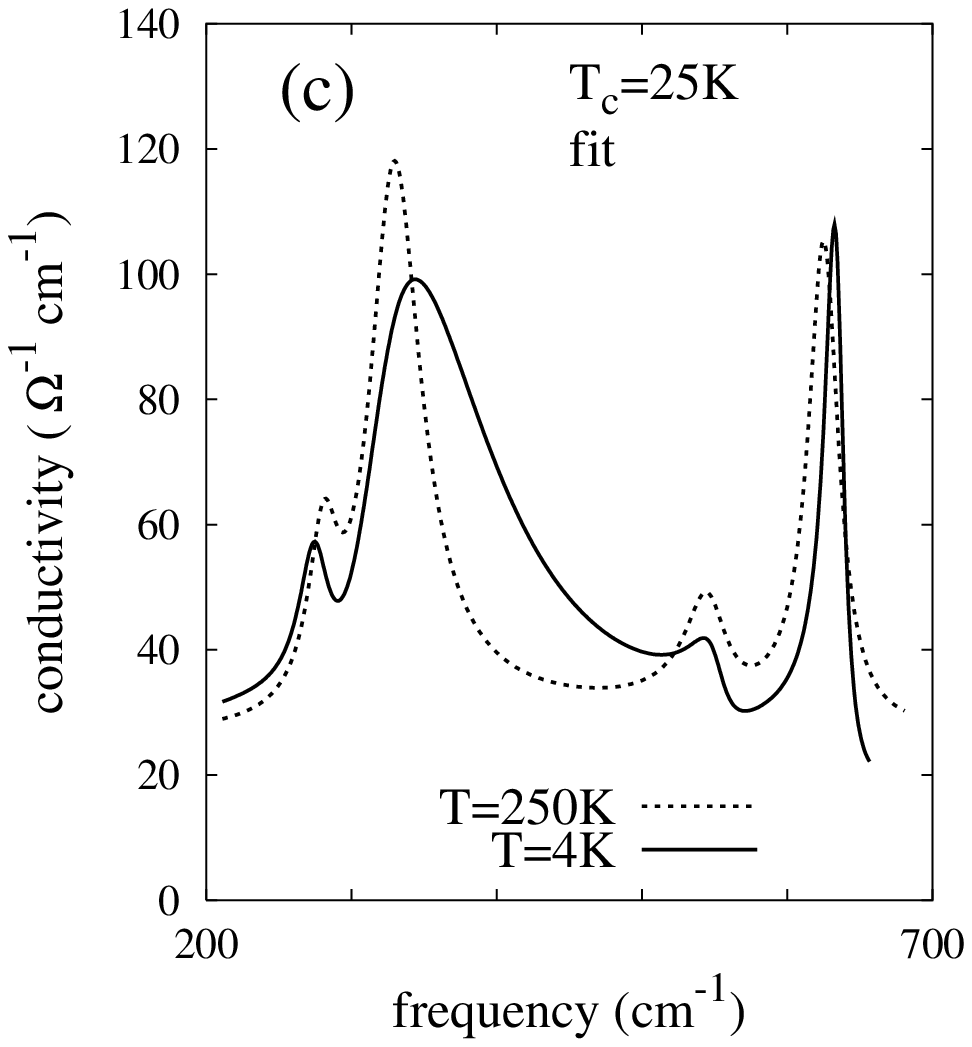} 
        \epsfysize = 4.4 cm
        \epsfbox{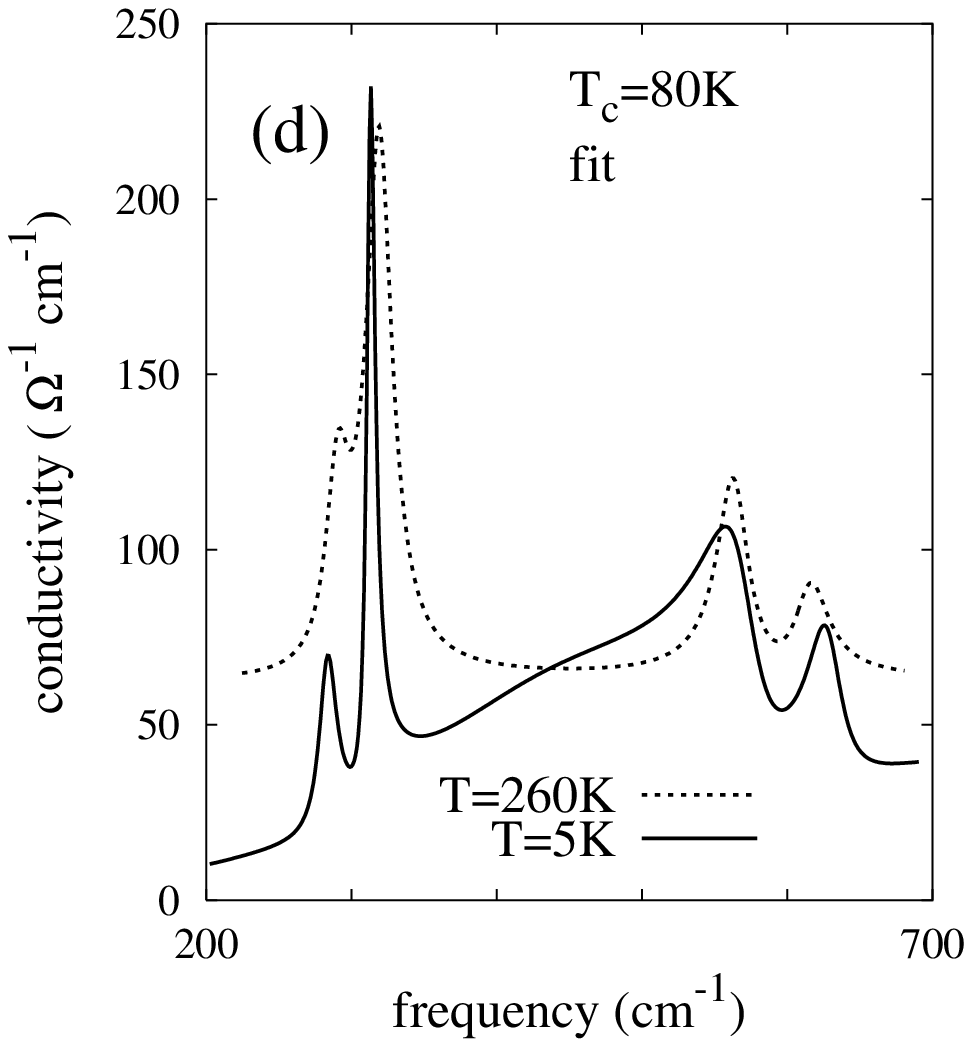} } }
\vskip 0.2 cm 
\caption{
Experimental spectra  
of the $c$-axis conductivity of YBa$_{2}$Cu$_{3}$O$_{y}$ 
with (a) $y=6.4$ ($T_{c}=25 {\rm\,K}$) and 
     (b) $y=6.8$ ($T_{c}=80 {\rm\,K}$)  
from Ref.~[15]. 
(c) and (d) Fits of these spectra obtained by using 
the present model. 
}
\end{figure}
The model allows us to explain not only 
the additional broad peak around $450 {\rm\,cm^{-1}}$ 
but also the spectacular anomaly  
of the oxygen bending mode at $320 {\rm\,cm^{-1}}$
and the spectral weight anomalies 
of the apical oxygen modes at $550$ 
and $630 {\rm\,cm^{-1}}$. 
Our results indicate  
that also the closely spaced copper-oxygen planes  
of underdoped bilayer cuprates 
are weakly (Josephson) coupled.  
These findings provide support for the conjecture \cite{Anderson}
that the $c$-axis dynamics of the cuprates 
(at least the underdoped ones) 
is dictated by the unconventional properties 
of the ground state of the electronic 
system of the planes. 
We suggest 
that the onset of the anomalies 
at temperatures significantly higher than $T_{c}$ 
may be caused 
by pairing fluctuations within  
the bilayers.  

We thank G.~P.~Williams and L.~Carr 
for technical support at the U4IR beamline at NSLS
and E.~Br{\"u}cher and R.~Kremer for SQUID measurements. 
We acknowledge discussions with 
M.~Gr{\"u}ninger,
D.~van der Marel, 
R.~Zeyher, T.~Strohm, and A.~Wittlin. 
D.~M.~gratefully acknowledges support by the 
Alexander von Humboldt Foundation. 
D.~M.~and J.~H.~were supported 
by the grant VS96102 of the Ministry of Education 
of the Czech Republic. 
\end{multicols}


\begin{thebibliography}{La86}
\bibitem[*]{address} 
Permanent address: Department of Solid State Physics
and Laboratory of Thin Films and Nanostructures,  
Faculty of Science, Masaryk University, Kotl{\'a}{\v r}sk{\'a} 2, 
CZ-61137 Brno, Czech Republic. 
\bibitem[**]{Aaddress} 
Permanent address: IFD, Warsaw University, Hoza, 69, 
PL-00-681 Warsaw, Poland. 
\bibitem{Rapp} M.~Rapp {\it et al.}, 
Phys.~Rev.~Lett.~{\bf 77}, 928 (1996). 
\bibitem{Schlenga} K.~Schlenga {\it et al.}, 
Phys.~Rev.~B {\bf 57}, 14 518 (1998) and references therein. 
\bibitem{Matsuda}  Y. Matsuda et al., 
Phys. Rev. Lett. {\bf 75}, 4512 (1995). 
\bibitem{Shibata} H.~Shibata, T.~Yamada, 
Phys.~Rev.~Lett {\bf 81}, 3519 (1998). 
\bibitem{Tamasaku}  K. Tamasaku, Y. Nakamura, and S. Uchida, 
Phys.~Rev.~Lett. {\bf 69}, 1455 (1992).
\bibitem{Bentum} P.~J.~M.~van Bentum {\it et al.}, 
Physica B {\bf 211}, 260 (1995). 
\bibitem{Anderson} P.~W.~Anderson, {\it The Theory of Superconductivity 
in the High-$T_{c}$ Cuprates}, 
Princeton University Press, Princeton, NJ, 1997. 
\bibitem{Emery} V.~J.~Emery, S.~A.~Kivelson, and O.~Zachar, 
Physica C {\bf 282-287}, 174 (1997) and references therein. 
\bibitem{Homes1} C.~C.~Homes {\it et al.}, 
Phys.~Rev.~Lett.~{\bf 71}, 1645 (1993). 
\bibitem{Homes2} C.~C.~Homes {\it et al.}, 
Physica C {\bf 254}, 265 (1995).  
\bibitem{Schutzman} J.~Sch{\"u}tzmann {\it et al.}, 
Phys.~Rev.~B {\bf 52}, 13 665 (1995). 
\bibitem{Hauff} R.~Hauff {\it et al.},  
Phys.~Rev.~Lett.~{\bf 77}, 4620 (1996). 
\bibitem{Tajima} S.~Tajima {\it et al.}, 
Phys.~Rev.~B {\bf 55}, 6051 (1997). 
\bibitem{Bernhard1} C.~Bernhard {\it et al.}, 
Phys.~Rev.~B {\bf 59}, 6631 (1999); 
C.~Bernhard {\it et al.}, 
Phys.~Rev.~Lett.~{\bf 80}, 1762 (1998). 
\bibitem{Bernhard2} C.~Bernhard {\it et al.}, unpublished. 
\bibitem{Zetterer} T.~Zetterer {\it et al.}, 
Phys.~Rev.~B {\bf 41}, 9499 (1990). 
\bibitem{Litvinchuk} A.~P.~Litvinchuk,
C.~Thompsen, M.~Cardona, 
Solid State Commun.~{\bf 83}, 343 (1992); 
A.~P.~Litvinchuk {\it et al.}, 
Z.~Phys.~B {\bf 92}, 9 (1993). 
\bibitem{Basov} D.~N.~Basov {\it et al.}, 
Phys.~Rev.~B {\bf 50}, 3511 (1994). 
\bibitem{Osafune} T.~Osafune {\it et al.}, 
Phys.~Rev.~Lett.~{\bf 82}, 1313 (1999). 
\bibitem{VdMarel2} D.~van der Marel and coworkers, private communication. 
\bibitem{VdMarel1} D.~Van der Marel, A.~Tsvetkov,   
Czech.~Journ.~of Phys.~{\bf 46}, 3165 (1996).  
\bibitem{Humlicek} J.~Huml{\'\i}{\v c}ek {\it et al.}, 
Physica C {\bf 206}, 345 (1993). 
\bibitem{Henn} R.~Henn {\it et al.}, 
Phys.~Rev.~B {\bf 55}, 3285 (1997). 
\bibitem{elst.}
We have represented the ions by uniformly charged planes 
perpendicular to the $c$-axis 
and expressed 
the additional electrical fields generated by the phonon displacements.  
The electrostatical calculations lead to expressions 
for the parameters $\alpha$, $\beta$, $\gamma$. 
The values of the charge densities of the planes 
have been estimated 
using the values of the effective ionic charges  
presented in Ref.~\cite{Henn}. 
In order to express the fields $E_{bl}$ and $E_{int}$,  
the boundaries between the intra- and inter-bilayer regions 
have to be specified. We have identified them  
with the charged planes corresponding to the oxygens.
The use of this model may be partially justified 
by the fact that the electromagnetic skin depth  
is shorter than the wavelength, 
so that any pronounced cancellation 
of the fields generated by the phonons 
is not expected. 
\bibitem{Basov2} D.~N.~Basov {\it et al.}, 
Science {\bf 283}, 49 (1999). 
\bibitem{Mihailovic} D.~Mihailovic, T.~Mertelj, K.~A.~M{\"u}ller, 
Phys.~Rev.~B {\bf 57}, 6116 (1998). 
\end{thebibliography}
\end{document}